# First-principles mode-by-mode analysis for electron-phonon scattering channels and mean free path spectra in GaAs

Te-Huan Liu[1], Jiawei Zhou[1], Bolin Liao[1], David J. Singh[2], and Gang Chen[1,*]

[1]Department of Mechanical Engineering, Massachusetts Institute of Technology, Cambridge, MA, 02139, USA

[2]Department of Physics and Astronomy, University of Missouri, Columbia, MO, 65211, USA

**Abstract**

We present a first-principles framework to investigate the electron scattering channels and transport properties for polar material by combining the exact solution of linearized electron-phonon (*e*-ph) Boltzmann transport equation in its integral-differential form associated with the *e*-ph coupling matrices obtained from polar Wannier interpolation scheme. No *ad hoc* parameter is required throughout this calculation, and GaAs, a well-studied polar material, is used as an example to demonstrate this method. In this work, the long-range and short-range contributions as well as the intravalley and intervalley transitions in the *e*-ph interactions (EPIs) have been quantitatively addressed. Promoted by such mode-by-mode analysis, we find that in GaAs, the piezoelectric scattering is comparable to deformation-potential scattering for electron scatterings by acoustic phonons in EPI even at room temperature and makes a significant contribution to mobility. Furthermore, we achieved good agreements with experimental data for the mobility, and identified that electrons with mean free paths between 130 and 210 nm contribute dominantly to the electron transport at 300 K. Such information provides deeper understandings on the electron transport in GaAs, and the presented framework can be readily applied to other polar materials.

---

[*]Author to whom correspondence should be addressed. E-mail: gchen2@mit.edu

# I. Introduction

Electron-phonon interaction (EPI) was first studied by Bloch [1], who discussed the interference between electrons and elastic waves and the effect on the temperature dependence of electrical conductivity. It was then realized that the EPI is in the center of the electron transport, being the key limiting factor of the carriers' lifetime in ordinary metals [2,3] and semiconductors [4-8] in a large temperature range, apart from its other roles in the hot-electron thermalization process [9,10] as well as the superconductivity [11,12]. The essence of EPI lies in the fact that, when the atoms move around due to thermal vibrations, the potential seen by electrons are disturbed. Such perturbation leads to exchange of energy and momentum between electrons and phonons, which determines the electron lifetimes. Based on this picture, quasi-analytical forms of electron-phonon ($e$-ph) coupling matrices for different types of scattering mechanisms have been derived, and are employed to obtain scattering rates based on Fermi's golden rule, either analytically or computationally, which are then used to explain electron transport properties [7,8]. Bardeen and Shockley [4] first proposed an insightful concept known as the deformation potential for the EPI in semiconductors, which relates the strength of electron-acoustic-phonon interaction to the shift of band edge due to local strain. Long-wavelength acoustic phonon generates a corresponding dilation of crystal, and the relative changes in atomic spacing gives rise to a perturbed potential that tends to shift the electronic band energy and to couple with electrons. In this case the perturbed potential is approximately proportional to strain, or $V_{e\text{-ph}}=\Xi_{ADP}\nabla\cdot\mathbf{u}$, where $\Xi_{ADP}$ is the acoustic deformation potential and $\mathbf{u}$ is the displacement of atom. Herring and Vogt [5] generalized the deformation-potential theory, and computed relaxation time tensors and mobilities of silicon and $n$-type germanium. Harrison [13] extended the deformation-potential theory to optical phonons that exists when each lattice point has two or more atoms as the basis.

For long-wavelength optical phonons, the atoms vibrate against each other without changing the size of unit cell in contrast to acoustic phonons, and the variations of the distance between basis atoms directly disturb the potential around the lattice, which acts a scattering source to electrons. In this picture the perturbed potential is nearly proportional to atomic displacement as $V_{e\text{-ph}}=\Xi_{ODP}u$ [14], where $\Xi_{ODP}$ is the optical deformation potential.

The deformation potential describes only the short-range interactions between electrons and long-wavelength phonons. When polar materials are considered, the electric dipole field generated by charged atoms decays slowly. The long-range interactions cause additional scattering for electrons—the long-wavelength acoustic phonon induces lattice strain which is called piezoelectric scattering, while the long-wavelength optical phonon induces bond stretching in lattice which is called polar-optical-phonon scattering. Meijer and Polder [15] first discussed the electron scattering by macroscopic piezoelectric fields which is generated by lattice distortion when the inversion center is absent in crystal. Their model study reveals that the piezoelectric scattering is predominate at very low temperatures. On the other hand, the longitudinal-optical (LO) phonon serves a strong scattering source to electron in polar materials due to the induced macroscopic polarization field, which was first investigated by Fröhlich [16]. If the dielectric screening is small, the direct effect of the dipole field can be dominant and therefore the polar-optical-phonon interaction is usually the major scattering channel in polar materials at room temperature. A good discussion on short-range and long-range EPIs is given by Vogl [17].

We summarize the semi-empirical formulae of $e$-ph coupling matrix as well as scattering rate for the four scattering mechanisms in table I; for polar-optical-phonon interaction, it is usually called Fröhlich model, while for piezoelectric interaction is called Hutson model [18]. These phenomenological models are widely employed to study the electron transport properties in non-polar materials through the Boltzmann transport equation (BTE) under relaxation time approximation (RTA) [5,10,13,18]. In polar semiconductors however, the validity of the RTA is questionable owing to the strong inelastic scattering caused by LO phonons [7,8]. Variational principle [19-21], Rode's iterative scheme [22-28] as well as Monte Carlo method [29-31] have been employed to extract the electron mobility of polar materials. However, these easily used semi-empirical models have strong limitations to qualitatively investigate the EPIs mode-by-mode due to the simplifications made in derivations. Models for the coupling matrices only depend on the magnitude of phonon wavevector; the periodic part of Bloch function near the band edge at which most transition events occur is usually changed smoothly, and therefore the overlap integral can be taken to be unit to take away the dependence of electron state [32]. This causes that such models are only applicable to isotropic crystals. The Debye model is used for acoustic phonons to describe the linear dispersion behavior at long-wavelength limit, while the dispersionless assumption is made for optical phonons. Furthermore, the parabolic band assumption is implemented to achieve the explicit formulae for electron scattering rate. To facilitate the understanding of materials and promote the discovery of new materials, it therefore becomes necessary to develop a method that can calculate the electrical properties without these assumptions.

The efforts toward such a goal have been evidenced by the development of the density-functional perturbation theory (DFPT) [33] and the Wannier interpolation scheme [34], which allows the determination of the *e*-ph coupling matrix fully from first-principles calculations [35]. Restrepo *et al*. [36] presented the first ab initio mobility for the case of silicon. These techniques then have been widely used to investigate the EPIs and to compute the thermoelectric properties [37-39] and electron mean free path (MFP) spectra [39] in silicon as well as the electrical resistivity in graphene [40-42] under RTA. Such density-functional-theory-based (DFT-based) treatment can also be employed to compute electron mobility in weakly polar materials such as transition metal dichalcogenides [43-46] and perovskites [47,48] that RTA still works due to the LO-phonon scatterings are suppressed by strong dielectric screening. For strongly polar materials, like GaAs, the long-range information originated from polar-optical-phonon and piezoelectric interactions in *e*-ph coupling matrices are lost during the Wannier interpolation [35]. The missing of such long-range contributions makes first-principles calculations of the scattering rate in polar semiconductors become more challenging, until recently when the "polar Wannier interpolation scheme" was proposed by Sjakste *et al.* [49], and by Verdi and Giustino [50], which adds in the long-range *e*-ph couplings. Despite these developments and the scattering rates obtained via such polar Wannier interpolation process, efforts to compute electron transport property such as mobility in polar materials by first-principle calculation is just beginning. In addition, the electron MFP spectra for polar semiconductors still remain unclear. The challenges result from, firstly, a very fine sampling of Brillouin zone is necessarily required to ensure the convergence of transport properties, which is computationally extremely demanding. Secondly, the RTA is not well justified in polar materials, and instead linearized BTE should be solved iteratively on a very fine mesh in order to determine the electron distribution function.

This paper conducts a detailed first-principle study of the electron transport property in GaAs, a prototypical polar semiconductor with well-documented properties. The coupling matrices are computed via the polar Wannier interpolation scheme, which are then used to obtain the electron scattering rates. The transport properties such as electron MFP are extracted from the exact solution of linearized BTE obtained by an iterative scheme. Although GaAs has been a well-known material, our mode-by-mode analysis enables us to quantitatively determine the contribution of long-range and short-range scatterings as well as of intravalley and intervalley transitions in EPI, which shines light on the detailed scattering mechanisms that differ from existing knowledge, for example, the trend of electron scattering rate especially close to Γ point and the important contribution from piezoelectric interaction even at room temperature. The obtained spectral distribution of electron MFP is also useful when designing nanodevices using GaAs. More importantly, we expect this method to be applied to other materials and help understand their electron transport behavior.

**II. Methodology**

The key of studying electron transport is the determination of the *e*-ph coupling matrix within the first-principles framework, and then compute the transport property such as electron mobility by the exact solution of linearized *e*-ph BTE. The *e*-ph coupling matrix is given by [6]

$$\mathbf{M}_{n\mathbf{k},p\mathbf{q}}^{m\mathbf{k}+\mathbf{q}} = \left(\frac{\hbar}{2m_0\omega_{p\mathbf{q}}}\right)^{1/2} \langle m\mathbf{k}+\mathbf{q}|\delta V_{p\mathbf{q}}(\mathbf{r})|n\mathbf{k}\rangle , \qquad (1)$$

where $m_0$ is a reference mass, $|n\mathbf{k}\rangle$ is the periodic part in electron wave function $\psi_{n\mathbf{k}}(\mathbf{r})=\langle \mathbf{r}|n\mathbf{k}\rangle$, and $n\mathbf{k}$ and $p\mathbf{q}$ represent the wavevector for electron at $n$ band and for phonon at $p$ mode respectively. $\delta V_{p\mathbf{q}}$ is the perturbed potential due to phonon vibration that can be computed by DFPT calculations. It should be emphasized that the DFPT calculation cannot yield a very precise value for piezoelectric coefficient. For the III-V group semiconductors, the averaged calculation error of the first-order piezoelectric constant is 10% [51], and therefore about a 20% error will be included in the piezoelectric interaction since the coupling strength is measured proportional to the square of piezoelectric constant.

In this section, we will show the derivation for the linearized *e*-ph BTE and its exact solution solved by means of the iterative method. This solution associated with the *e*-ph coupling matrix obtained by polar Wannier interpolation will be used to compute the electron scattering rates as well as the transport properties. The simulation details and the convergence tests are also presented.

**A. Iterative electron-phonon Boltzmann transport equation**

Assuming the absence of temperature and electrochemical gradients, a system at steady state the BTE can be written as [52]

$$\left(\frac{\partial f_{n\mathbf{k}}}{\partial t}\right)_{coll} = -\left(\frac{\partial f_{n\mathbf{k}}}{\partial t}\right)_{diff}. \tag{2}$$

In Eq. (2), the diffusion term on the right-hand side is given by $\left(\partial f_{n\mathbf{k}}/\partial t\right)_{diff} = -e\mathbf{E}\cdot\mathbf{v}_{n\mathbf{k}}\,\partial f_{n\mathbf{k}}/\partial\varepsilon_{n\mathbf{k}}$, where $f_{n\mathbf{k}}$ is the distribution function of electron in *n*-th band at wavevector **k**. $\varepsilon_{n\mathbf{k}}$ and $\mathbf{v}_{n\mathbf{k}}$ are the electron energy and group velocity, respectively. $e\mathbf{E}$ is the electrical force acting on electron. If we fix our attention to a specific state *n***k**, an electron can be scattered into or out from that by absorbing or emitting a phonon of state ±*p***q**. In this view, there are four mechanisms that should be taken into account in the three-carrier interaction. In terms of the first-order perturbation theory using Fermi's golden rule, the collision term in Eq. (2) can be evaluated by [52],

$$\left(\frac{\partial f_{n\mathbf{k}}}{\partial t}\right)_{coll} = \frac{2\pi}{\hbar}\sum_{m\mathbf{k+q}}\left|M^{m\mathbf{k+q}}_{n\mathbf{k},p\mathbf{q}}\right|^2 \left\{\begin{bmatrix}(f^0_{n\mathbf{k}}+n^0_{-p\mathbf{q}})f'_{m\mathbf{k+q}}\\-(1+n^0_{-p\mathbf{q}}-f^0_{m\mathbf{k+q}})f'_{n\mathbf{k}}\end{bmatrix}\delta(\varepsilon_{n\mathbf{k}}-\varepsilon_{m\mathbf{k+q}}-\hbar\omega_{-p\mathbf{q}}) \\ +\begin{bmatrix}(1-f^0_{n\mathbf{k}}+n^0_{p\mathbf{q}})f'_{m\mathbf{k+q}}\\-(n^0_{p\mathbf{q}}+f^0_{m\mathbf{k+q}})f'_{n\mathbf{k}}\end{bmatrix}\delta(\varepsilon_{n\mathbf{k}}+\hbar\omega_{p\mathbf{q}}-\varepsilon_{m\mathbf{k+q}})\end{array}\right\}, \tag{3}$$

where $\hbar\omega_{p\mathbf{q}}$ is energy of phonon in $p$-th mode at wavevector $\mathbf{q}$, respectively. Equation (3) is derived based on the energy conservation conditions $(1-f_{n\mathbf{k}}^0)n_{-p\mathbf{q}}^0 f_{m\mathbf{k}+\mathbf{q}}^0 = f_{n\mathbf{k}}^0(1+n_{-p\mathbf{q}}^0)(1-f_{m\mathbf{k}+\mathbf{q}}^0)$ and $(1-f_{n\mathbf{k}}^0)(1+n_{p\mathbf{q}}^0)f_{m\mathbf{k}+\mathbf{q}}^0 = f_{n\mathbf{k}}^0 n_{p\mathbf{q}}^0(1-f_{m\mathbf{k}+\mathbf{q}}^0)$, and the electron distribution function is already written in terms of the deviation from the equilibrium distribution as $f_{n\mathbf{k}} = f_{n\mathbf{k}}^0 + f_{n\mathbf{k}}'$. In this work, the phonons are always assumed to be at their equilibrium states, essentially neglecting the phonon drag effect that mainly occurs at low temperature in semiconductors [53,54].

In low-field transport regime, the deviation of distribution function can be treated as a small perturbation from equilibrium distribution. Based on this assumption, the diffusion term in Eq. (2) can then be rewritten as $(\partial f_{n\mathbf{k}}'/\partial t)_{diff} = -e\mathbf{E}\cdot\mathbf{v}_{n\mathbf{k}}\,\partial f_{n\mathbf{k}}^0/\partial\varepsilon_{n\mathbf{k}}$. The single-mode RTA claims that $f_{n\mathbf{k}}'$ should vanish after a time period $\tau_{n\mathbf{k}}$, and thereby the deviation of distribution function can be obtained explicitly, which reads $f_{n\mathbf{k}}' = e\mathbf{E}\cdot\mathbf{v}_{n\mathbf{k}}\tau_{n\mathbf{k}}\,\partial f_{n\mathbf{k}}^0/\partial\varepsilon_{n\mathbf{k}}$. Here we can define a electron mean free displacement $\mathbf{F}_{n\mathbf{k}} \equiv \mathbf{v}_{n\mathbf{k}}\tau_{n\mathbf{k}}$, which should be determined by iterative procedure as will be shown below, and the deviation of distribution function can be rewritten as

$$f_{n\mathbf{k}}' = -\frac{f_{n\mathbf{k}}^0(1-f_{n\mathbf{k}}^0)}{k_B T}e\mathbf{E}\cdot\mathbf{F}_{n\mathbf{k}}. \qquad (4)$$

We make use of Eq. (4) to linearize the collision integral in Eq. (3) and the diffusion term in Eq. (2). After equating Eqs. (2) and (3), the BTE is given by

$$\mathbf{v}_{n\mathbf{k}} = \sum_{m\mathbf{k}+\mathbf{q}}\left(G_{n\mathbf{k}}^{m\mathbf{k}+\mathbf{q},-p\mathbf{q}} + G_{n\mathbf{k},p\mathbf{q}}^{m\mathbf{k}+\mathbf{q}}\right)\left(\mathbf{F}_{n\mathbf{k}} - \mathbf{F}_{m\mathbf{k}+\mathbf{q}}\right), \qquad (5)$$

where $G_{n\mathbf{k}}^{m\mathbf{k+q},-p\mathbf{q}} = 2\pi/\hbar \left|M_{n\mathbf{k},p\mathbf{q}}^{m\mathbf{k+q}}\right|^2 \left(1+n_{-p\mathbf{q}}^0 - f_{m\mathbf{k+q}}^0\right) \delta(\varepsilon_{n\mathbf{k}} - \varepsilon_{m\mathbf{k+q}} - \hbar\omega_{-p\mathbf{q}})$ and $G_{n\mathbf{k},p\mathbf{q}}^{m\mathbf{k+q}} = 2\pi/\hbar \left|M_{n\mathbf{k},p\mathbf{q}}^{m\mathbf{k+q}}\right|^2 \left(n_{p\mathbf{q}}^0 + f_{m\mathbf{k+q}}^0\right) \delta(\varepsilon_{n\mathbf{k}} + \hbar\omega_{p\mathbf{q}} - \varepsilon_{m\mathbf{k+q}})$ are the electron transition rates at equilibrium due to the phonon emission and absorption processes, respectively, which should be computed by the first-principles calculations in this work. Equation (5) can be further rewritten as an iterative format:

$$\mathbf{F}_{n\mathbf{k}}^{(i+1)} = \tau_{n\mathbf{k}}^{\mathrm{RTA}} \left[\mathbf{v}_{n\mathbf{k}} + \sum_{m\mathbf{k+q}} \left(G_{n\mathbf{k}}^{m\mathbf{k+q},-p\mathbf{q}} + G_{n\mathbf{k},p\mathbf{q}}^{m\mathbf{k+q}}\right) \mathbf{F}_{m\mathbf{k+q}}^{(i)}\right], \quad (6)$$

which is so-called iterative e-ph BTE. Here $\tau_{n\mathbf{k}}^{\mathrm{RTA}} = \left[\sum_{m\mathbf{k+q}} \left(G_{n\mathbf{k}}^{m\mathbf{k+q},-p\mathbf{q}} + G_{n\mathbf{k},p\mathbf{q}}^{m\mathbf{k+q}}\right)\right]^{-1}$ is the relaxation time of electron according to the Migdal approximation [55]. In this work, we deal with the surface integrals in the relaxation time by means of tetrahedral integration [56]. Once the relaxation time is determined, the mean free displacement $\mathbf{F}_{n\mathbf{k}}$ can therefore be solved iteratively. It is straightforwardly to set the initial $\mathbf{F}_{m\mathbf{k+q}}^{(0)}$ to be zero, and thus the first-order solution $\mathbf{F}_{n\mathbf{k}}^{(1)} = \mathbf{v}_{n\mathbf{k}} \tau_{n\mathbf{k}}^{\mathrm{RTA}}$ is naturally the MFP of electron under RTA. And then, $\mathbf{F}_{m\mathbf{k+q}}^{(1)}$ can be obtained by $\mathbf{F}_{n\mathbf{k}}^{(1)}$ with a difference of phonon wavevector $\mathbf{q}$ to carry on the self-consistent calculations. The iteration procedure provides us with the true deviation of electron distribution function given by Eq. (4) in the low-field transport for polar materials. Once the converged deviation function is achieved, the electron mobility tensor can be calculated by

$$\mu_{\alpha\beta} = \frac{2e}{\Omega N_{\mathbf{k}} n_c} \sum_{n\mathbf{k}} v_{n\mathbf{k},\alpha} F_{n\mathbf{k},\beta} \frac{\partial f_{n\mathbf{k}}^0}{\partial \varepsilon_{n\mathbf{k}}}, \quad (7)$$

where $n_c$ is the carrier concentration, and $N_{\mathbf{k}}$ is the number of k-point grids. The subscripts $\alpha$ and $\beta$ denote the direction of electron transport.

## B. Simulation details and convergence test

In the very first step, there are three physical quantities that should be determined: electronic Hamiltonian, dynamical matrix and phonon perturbation. These calculations are performed on the Quantum ESPRESSO package [57]. A Troullier-Martins norm-conserving pseudopotential with the Perdew-Wang exchange-correlation functional is employed to describe the interactions of Ga and As atoms. The cutoff energy of plane wave is chosen as 80 Ry, and the 20×20×20 and 6×6×6 Monkhorst-Pack k-point meshes are used for the self-consistent and non-self-consistent field calculations, respectively. The convergence threshold of energy is set to be $10^{-12}$ Ry. The obtained optimized lattice constant for GaAs is 5.53 Å. In the phonon calculations, the dynamical matrices and phonon perturbations are computed on a 6×6×6 q-point mesh. We suggest the threshold of phonon calculations being $10^{-22}$ Ry for the better convergence of the phonon perturbation. We use the BerkeleyGW package [58] to achieve a more accurate electron band structure. The cutoff energies of the screened and bare Coulomb potential are set by 15 and 45 Ry, respectively. 80 empty conduction bands are left in order to calculate the dielectric constant and Green's function. The obtained bandgap is 1.411 eV, and the energy differences between Γ and L, and Γ and X valleys are 0.258 and 0.387 eV, respectively. The GW eigenvalues are computed on a 6×6×6 k-point mesh as the input in the subsequent calculations.

In this work, the EPW [59] package is employed to interpolate the *e*-ph coupling matrices as well as the electron and phonon eigenvalues from the coarse (obtained by DFT and DFPT calculations as mentioned) to fine k- and q-point meshes using polar Wannier interpolation. Our in-house code, an iterative solver for the linearized *e*-ph BTE, is used to calculate the electron transport properties. Figures 1(a) and (b) show the convergence of electron mobility with respect to different q- and k-point grids, respectively. We found that the calculation on 600×600×600 k-point mesh associated with 100×100×100 q-point mesh achieves the best balance between the convergence of mobility and the computational efficiency, and therefore this mesh will be used for presenting transport properties throughout this work. The mobility is considered convergent while the variation is less than 0.1% as shown in the Fig. 1(c). The carrier concentration is chosen as $10^{13}$ cm$^{-3}$ to characterize the Fermi level of an intrinsic GaAs. In this case, at 300 K, we found that the mobility has about 18% increases compared to RTA's result once the convergence of the solution of *e*-ph BTE is achieved. Figure 1(d) demonstrates the electron scattering rates near Γ point before and after iteration. The scattering rate during the iterations is defined by the inverse of effective relaxation time, $\tau^{-1}_{eff,n\mathbf{k}}=(\mathbf{F}_{n\mathbf{k}}\cdot\mathbf{v}_{n\mathbf{k}}/|\mathbf{v}_{n\mathbf{k}}|^2)^{-1}$. The significant decreasing of scattering rates shows the importance to solve linearized BTE iteratively in studying electron transport properties in polar materials.

## III. Results and discussions

In Fig. 2, we demonstrate the capability of first-principles calculation of electron scattering rates in GaAs. Figure 2(a) shows the energy-dependent scattering rate within 0.5 eV of conduction band edge. By comparing the first-principles results with the semi-empirical model, several significant differences can be seen. Our simulation shows that the scattering rate has a jump very close to the Γ point, which is due to the contribution of longitudinal-acoustic-phonon (LA-phonon) and transverse-acoustic-phonon (TA-phonon) scatterings, whereas the model largely underestimates the scatterings of acoustic phonons near the Γ point, and as a result the trend is still determined by the LO phonons. Furthermore, when the energy increases, after the phonon emission process starts to participate in the EPI, the computed scattering rate shows a slightly increasing trend versus energy at Γ valley that is in contrast to the model's prediction, which can be attributed to the non-parabolicity conduction band of GaAs and was discussed before [8]. In the higher-energy valleys, the first-principles calculations also shows remarkable inconsistency compared with the models, particularly at X valley. These discrepancies mainly come from the lack of consideration of the intervalley scattering in the semi-empirical model [8]. Overall, the mobility evaluated from model is 4930 cm$^2$/Vs at 300 K, while the first-principles calculation presents a more reasonable value, 7050 cm$^2$/Vs under RTA. These two effects—non-parabolicity band and intervalley scattering—are automatically included in the first-principles calculation, but are difficult to incorporate otherwise especially when modeling a complex band structure.

With the first-principles approach, as displayed in Fig. 2(b), we can now look at the EPI contribution from each phonon mode with respect to different energies and valleys. For polar materials, the momentum change of electrons majorly results from the scattering by LO phonons, which has been widely studied in literatures [6-8]. It also offers us a route to examine the detailed EPIs due to phonon absorption and emission processes mode-by-mode as shown in Fig. 2(c). The electron-LO-phonon interaction in semiconductor is dominated by the absorption process when the electron energy is smaller than a LO-phonon energy at zone center, $\hbar\omega_{LO,center}$, and then the LO-phonon emission process takes over the EPI as the electron energy increases. The scattering rates computed at 200, 300 and 400 K are shown in Fig. 2(d). We can see the scattering rates shift upward while the temperature increases due to the increased phonon population of each mode under the RTA (phonons are kept in their equilibrium states), consistent with theoretical prediction [60].

When we examine the *e*-ph coupling strength as displayed in Fig 3(a), except for the case of LO phonon which has been discussed before [49,50], it is found that the LA-phonon coupling strength $|M_q|^2$ also has remarkable increase after the polar Wannier interpolation scheme is applied. The obtained trend of coupling strength at long-wavelength phonon is proportional to $|\mathbf{q}|^{-1}$, which is in accordance with the semi-empirical model for piezoelectric interaction as listed in table I. This fact can be attributed to the fact that most of the long-range piezoelectric interactions are lost during the conventional interpolated procedure, while can be preserved during the polar Wannier interpolation, and also implies that the LA-phonon scattering rate might be largely underestimated in GaAs once the contribution of piezoelectric interaction is simply neglected. Figure 3(a) shows the electron scattering rates due to LA phonons calculated by using polar Wannier and Wannier interpolation as well as the differences between the two schemes. Generally, for the acoustic mode, one obtains $\varepsilon^{0.5}$ energy dependence for scattering rate from deformation-potential scattering, and this is well captured by Wannier interpolation. However, the calculation results obtained through the polar Wannier interpolation indicate that the energy dependence of LA-phonon scatterings is much different from that of the deformation-potential scattering. The scattering rates have at least a twofold increase within the studied energy region. By comparing the scattering rate between the two methods, we find that their difference can be well fitted by the typical trend of piezoelectric scattering as shown in table I. A theoretical study presented by Wolfe *et al*. [61] shows that in GaAs, the piezoelectric scattering is comparable to the deformation-potential scattering only when the temperature is lower than 100 K. However, our first-principles results indicate that the piezoelectric interaction dominates the electron-acoustic-phonon interaction in GaAs even at the room temperature. This EPI is inherently taken into account in DFPT (linear-response theory) calculation, which has been employed to compute the proper piezoelectric tensor [51,62]. Although DFPT can only provide the first-order contribution to the piezoelectric field, recent experimental and theoretical studies [63,64] show that the second-order effect in $In_xGa_{1-x}As$ is safely negligible when the concentration of In atom is low, that is, when the lattice strain is very small. It is worthy of noting that for both piezoelectric and deformation-potential scatterings, the trends fitted by models show remarkable inconsistencies with the first-principles results in GaAs, when the energy is lower than a LA-phonon energy at zone boundary, $\hbar\omega_{\text{LA,boundary}}$. This is because the models are derived based on the elastic assumption, which treats the phonon absorption and emission in the same way, and therefore cannot well characterize the scattering rate for emission processes at bottom of valley.

After discussing the *e*-ph scattering channels, we proceed to discuss the electron transport properties. The scattering rates displayed in Figs. 2 and 3 are computed based on the RTA. These, however, cannot be directly used to evaluate the transport properties in polar materials. Instead, the fully iterative scheme for the linearized *e*-ph BTE should be applied to correctly determine the deviation of distribution function from Eq. (4). This is expected to be particularly important for high mobility materials, whose electron transport is often dominated by long-range polar interaction. As displayed in Fig. 4, our calculation results are in good agreement with the experimental measurements in the range from 200 to 400 K [24,65-67]; the computed electron mobility at 300 K is 8340 cm$^2$/Vs. The mobilities obtained from the iterative method are always larger than that from the RTA in the temperature range we studied as indicated by the blue-dash line in Fig. 4. There are two perspectives on why RTA underestimates mobility in polar materials. One is from the phonon energy—the LO phonon has a non-negligible energy compared with electrons that are close to the conduction band edge, and as a consequence the electron-LO-phonon scattering cannot be considered as an elastic process [8]. Another view focuses on the magnitude of electron-LO-phonon coupling strength. The probability of electron interacting with a LO phonon with long wavelength is so high (see $\mathbf{M}_\mathbf{q}^{\text{POP}}$ in table I) that the distribution of the scattered electron can no longer be treated as close to the equilibrium state as claimed by the RTA [7]. Both perspectives point out a fact, that is in such cases relaxation time cannot be defined. Our simulations show that at very low temperature, there is no change in electron mobility after iteration since the LO phonons are not populated. As temperature increases, the electron-LO-phonon scattering starts to take over the normal EPI, and as a result the deviation of distribution function of scattered electron becomes larger. The calculation shows that at 200, 300 and 400 K, the iteration scheme, respectively, provides 11%, 18% and 22% correction to mobility compared with RTA, which mainly results from the accumulation of deviation of electron distribution function due to the strong coupling between electrons and long-wavelength phonons. The first-principles calculations can be used to more precisely quantify the effect of scatterings from different phonon modes on the electron transport. We have shown above that for the LA phonon both piezoelectric and deformation-potential scatterings contribute to the EPI near the room temperature. Here we further show that it has a non-negligible effect on the electron transport as displayed by the green-dash line in the inset in Fig. 4. The mobility increases by 23% to 12% when the temperature varies from 200 to 400 K, if we artificially exclude the electron-LA-phonon interactions. The trend also indicates that the acoustic-phonon scatterings become dominant at low temperatures, which is in accordance with the predictions by piezoelectric and deformation-potential theory.

Now we discuss another important property, the electron MFP. The mean free displacement $\mathbf{F}_{n\mathbf{k}}$ shown in Eq. (6) cannot characterize the MFP of the electron at eigenstate $n\mathbf{k}$ (it can be if only we put all $\mathbf{F}_{m\mathbf{k}+\mathbf{q}}$ as zero), due to the fact that it is mixed with other eigenstates $m\mathbf{k}+\mathbf{q}$ after iteration. Although the direction of the mean free displacement is no longer identical with energy flow, we can define an effective scalar MFP by projecting $\mathbf{F}_{n\mathbf{k}}$ to its group velocity as $\Lambda_{n\mathbf{k}}=\mathbf{F}_{n\mathbf{k}}\cdot\mathbf{v}_{n\mathbf{k}}/|\mathbf{v}_{n\mathbf{k}}|$ [68]. The MFP spectra at different temperatures shown in Fig. 5(a) is a crucial

quantity in engineering or designing nanostructure for various devices, but for GaAs (even for other polar semiconductors), it still remains unclear. Electron MFP of GaAs characterized by the Drude model is about 35 nm, which is obviously too small since the free electron approximation is unable to reflect the correct picture of an electron moving in a periodic potential. The presented MFP spectra show that electrons that contribute to the mobility have a narrow span of MFP, between 130 to 210 nm at 300 K. It also shows that the MFP spectra becomes wider at lower temperature since the dominant electron-LO-phonon scattering rate drops rapidly with the decrease of temperature, giving rise to an overall increase to the MFPs. The distribution of MFP with different energies is demonstrated in Fig. 5(b). By comparing Figs. 5(a) and (b), we can determine the energy range that contributes most to the mobility in nanostructures. In polar materials Howarth and Sondheimer [19] proposed that the MFP, as a function of energy and temperature, should be proportional to $\varepsilon T^{-1}$. The first-principles MFPs show the agreement with the trend after phonon emission process fully takes over the EPI (approximately when $\varepsilon_{n\mathbf{k}}$ is greater than 0.06 eV). However, near the conduction band edge, we can see the MFP has a local maximum takes place around the energy of $\hbar\omega_{\text{LO,center}}$ since only phonon absorption process have substantial contribution to the EPI, and then goes to zero as the group velocity is zero at the bottom of valley.

Finally, we want to discuss the electron's intravalley and intervalley transitions at L and X points, which reveal crucial information for understanding hot electron thermalization process, and is of great value in studying III-V semiconductor devices. The intervalley transitions due to deformation-potential scatterings in GaAs and other zinc-blende crystals have been studied by Zollner *et al.* [69,70]. Figure 6 shows the electron transition rates in color map at each participation phonon mode, where the completed EPIs have been taken into account in the first-principles calculation automatically. It displays the possible transition events of the EPI at the L and X points, and the dark-blue region means no scattering can be induced by these phonons. We do not show the case at Γ point, because most the scatterings at Γ point are induced by small-wavevector LO phonons due to its isolation from the other two valleys; the energy differences between Γ and L points and Γ and X points are about 0.258 and 0.387 eV, respectively. This indicates that the EPI is dominated by the electron-LO-phonon scatterings, and that the excited electron near the Γ point will thermalize towards the conduction band edge mainly by emitting long-wavelength LO phonons. On the contrary, the intervalley transitions have substantial contributions to the EPI at L and X points. In the case of electron at L and X points, we can see distinct regions where the transition usually happens, which can be denoted by the three characteristic phonon wavevectors: **q**<0.30 (intravalley), **q**~0.87 (intervalley) and **q**~1.00 (intervalley). For L point, **q**~0.87 represents the intervalley transition from L point to Γ and X valleys, and **q**~1.00 represents the intervalley transition from L point to the other degenerate L valleys. On the other hand, for X point, **q**~0.87 represents the transition from X point to L valley, and **q**~1.00 represents the intervalley transition from X point to Γ and to other degenerate X valleys. Our calculations indicate that the intervalley transitions contribute 13.2% and 57.6% to the total scattering rates of electrons at the L and X points, respectively. On the other hand, the contribution from acoustic phonons in EPI is 10.7% and 51.2% at L and X points, respectively. These facts demonstrate that at X point, the electron scattering is dominated by the intervalley transitions, and the contribution from acoustic phonons, particularly the LA mode, in the EPI cannot simply be neglected.

**IV. Conclusion**

In summary, we present a parameter-free first-principles framework to study the EPIs in GaAs. This general computational scheme further enables us to examine the details of each scattering event, which happens, particularly, around the bottom of the valleys. The computed electron mobilities are in good agreement with experimental results; however, detailed scattering mechanisms differ from semi-empirical models which are derived based upon a lot of assumptions. Furthermore, the mode-by-mode analysis enables us to identify the (i) *Importance of piezoelectric scatterings in polar materials*. The scattering rates of LA phonon have at least twofold increase when the long-range piezoelectric interaction can be well addressed by polar Wannier scheme. Both the piezoelectric and deformation potential scatterings of LA phonons have non-negligible contributions to the EPI even at room temperature; they account for 15% reduction of the mobility at 300 K and their presence changes the energy dependence of the scattering rate near band edge. (ii) *Spectral distribution of electron MFP*. The mobility is mainly contributed from electrons with MFPs between 130 and 210 nm at room temperature. The MFPs near the $\Gamma$ point are also presented, which is hard to be predicted by the proposed model. (iii) *Quantitative determination of intravalley and intervalley transitions*. At X point, the intervalley transitions provide 57.6% in the total *e*-ph scatterings, which is comparable to the intravalley transitions. We also identify that 51.2% scattering is contributed by acoustic phonons. Thus detailed information about specific channels for *e*-ph scattering can be obtained and used in designing materials as well as devices where phonon scattering including polar optical phonon scattering controls the mobility. We expect the presented general framework can be applied to other structures and facilitate the understanding of electron transport in new materials.

**Acknowledgements**

Te-Huan Liu and Jiawei Zhou contributed equally to this work. We thank Wei-Chun Hsu, Qichen Song and Zhiwei Ding for the helpful discussions. This article was supported by S$^3$TEC, an Energy Frontier Research Center funded by the United States Department of Energy, Office of Basic Energy Sciences, under Award DE-FG02-09ER46577 (for fundamental research on electron-phonon interaction in thermoelectric materials) and by DARPA MATRIX program HR0011-16-2-0041 (for developing and applying the simulation codes).

**Table captions**

TABLE I. Semi-empirical formulae for *e*-ph coupling matrix and scattering rate of different EPIs [8,32]. The abbreviations stand for, sequentially, acoustic-deformation-potential, optical-deformation-potential, piezoelectric and polar-optical-phonon scatterings. These formulae are particularly used to indicate the dependence of phonon wavevector $\mathbf{q}$ and electron energy $\varepsilon_\mathbf{k}$ of coupling matrix and scattering rate, respectively. Here $e$ is the electron charge, $m_e$ is the electron elective mass, $\Omega$ is the volume of unit cell, and $\rho$ is the mass density of material. $v_\mathbf{q}$, $\omega_\mathbf{q}$ and $n_\mathbf{q}^0$ are the phonon group velocity, frequency and distribution function at equilibrium state, respectively. The screening effect is shown by $\epsilon_\infty$ and $\epsilon_0$, which are the high-frequency (no lattice response) and static (including lattice response) dielectric constants respectively. In piezoelectric mechanism, $e_{\mathrm{PZ}}$ represents the first-order piezoelectric constant, and $\lambda_\mathrm{D}$ is the Debye screening length.

| Mechanism | Coupling matrix $\mathbf{M_q}$ | Scattering rate $\tau^{-1}(\varepsilon_\mathbf{k})$ |
|---|---|---|
| ADP | $\sqrt{\dfrac{\Xi_{\mathrm{ADP}}^2 \hbar |\mathbf{q}|}{2\rho\Omega v_\mathbf{q}}}$ | $\dfrac{\Xi_{\mathrm{ADP}}^2 (2m_e)^{3/2} k_\mathrm{B} T}{2\pi\hbar^3 \rho v_\mathbf{q}^2} \sqrt{\varepsilon_\mathbf{k}}$ |
| ODP | $\sqrt{\dfrac{\Xi_{\mathrm{ODP}}^2 \hbar}{2\rho\Omega \omega_\mathbf{q}}}$ | $\dfrac{\Xi_{\mathrm{ODP}}^2 (2m_e)^{3/2}}{4\pi\hbar^3 \rho \omega_\mathbf{q}} \left[\left(n_\mathbf{q}^0 + \dfrac{1}{2} \mp \dfrac{1}{2}\right)\sqrt{\varepsilon_\mathbf{k} \pm \hbar\omega_\mathbf{q}}\right]$ |
| PZ | $\sqrt{\dfrac{e_{\mathrm{PZ}}^2 \hbar}{2\rho\Omega v_\mathbf{q}} \dfrac{e^2}{\epsilon_\infty |\mathbf{q}|}}$ | $\dfrac{e_{\mathrm{PZ}}^2 e^2 k_\mathrm{B} T}{\pi\hbar^2 \epsilon_\infty^2 \sqrt{2\varepsilon_\mathbf{k}/m_e}} \ln\left(1 + \dfrac{8 m_e \varepsilon_\mathbf{k}/\hbar^2}{\lambda_\mathrm{D}^2}\right)$ |
| POP | $\dfrac{ie}{|\mathbf{q}|}\sqrt{\dfrac{2\pi\hbar\omega_\mathbf{q}}{\Omega}\left(\dfrac{1}{\epsilon_\infty} - \dfrac{1}{\epsilon_0}\right)}$ | $\dfrac{e^2 \omega_\mathbf{q} \left(\dfrac{1}{\epsilon_\infty} - \dfrac{1}{\epsilon_0}\right)}{2\pi\hbar\epsilon_\infty \sqrt{2\varepsilon_\mathbf{k}/m_e}} \left[\left(n_\mathbf{q}^0 + \dfrac{1}{2} \mp \dfrac{1}{2}\right)\sinh^{-1}\sqrt{\dfrac{\varepsilon_\mathbf{k}}{\hbar\omega_\mathbf{q}} - \dfrac{1}{2} \pm \dfrac{1}{2}}\right] |

**Figure captions**

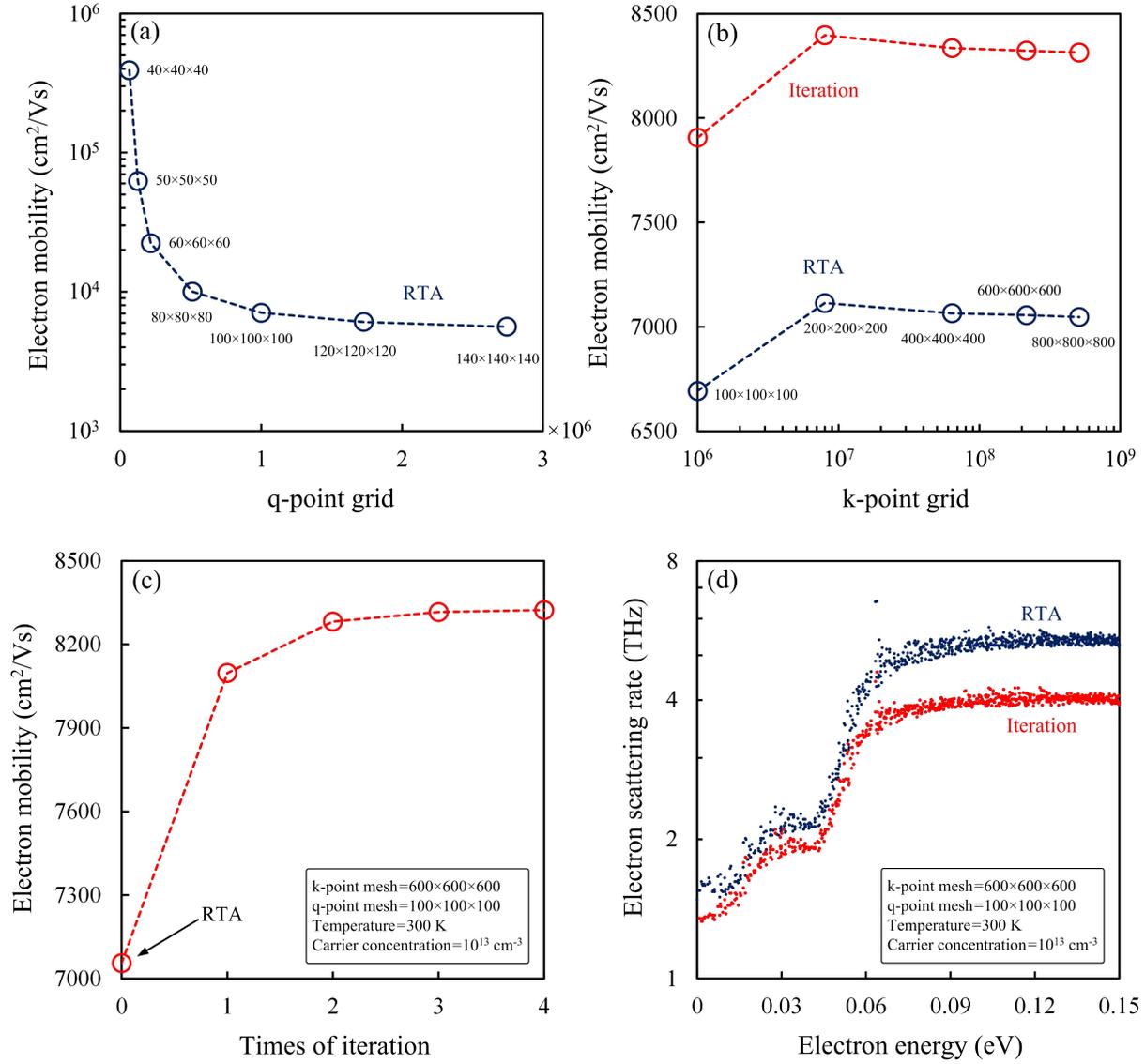

FIG. 1. (a) Electron mobility computed by a fixed 600×600×600 k-point mesh with respect to different number of q-point grids. (b) Electron mobility computed by a fixed 100×100×100 q-point mesh with respect to different number of k-point grids. The blue and red circles are the mobilities calculated from RTA and iterative scheme, respectively. (c) Evolution of electron mobility with respect to times of iteration. (d) Electron scattering rates near Γ point. The blue and red dots are the scattering rates computed by using RTA and iterative scheme, respectively,

from the 600×600×600 k-point and 100×100×100 q-point meshes. The studied temperature and carrier concentration in these simulations as shown in the four figures are 300 K and $10^{13}$ cm$^{-3}$, respectively.

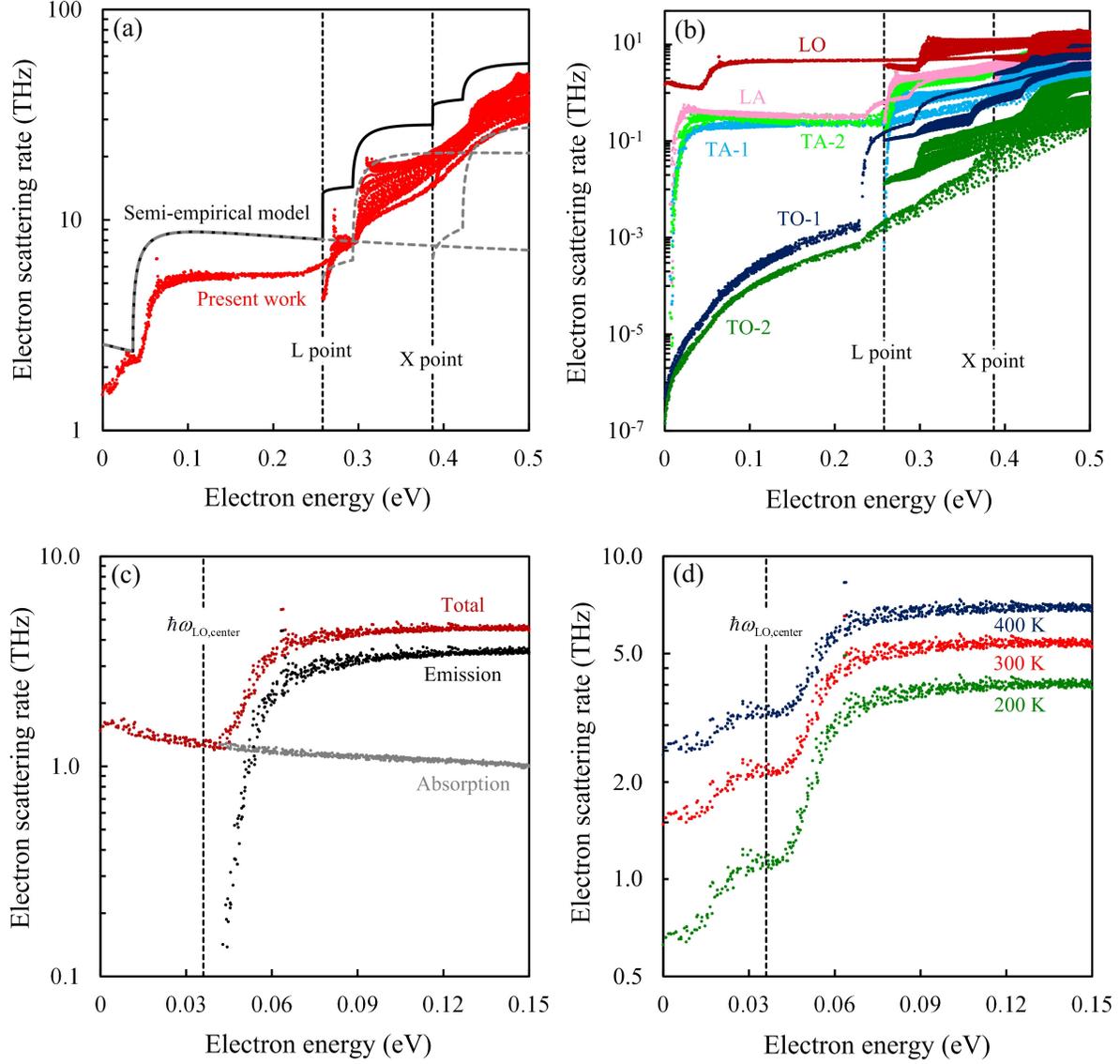

FIG. 2. (a) Electron scattering rate with respect to different electron energy at 300 K. The red dots are computed by present first-principles calculation. The black (total) and gray-dash (each valley) lines are predicted by the semi-empirical models including acoustic-deformation-potential, piezoelectric and polar-optical-phonon scatterings, which are listed in table I. The parameters in these formulae are provided by Ref. [8] (b) Electron scattering rate with respect to different electron energy due to each phonon mode at 300 K. (c) LO-phonon-limited electron scattering rate near Γ point at 300 K. The black and gray dots are the scattering rates due to LO-

phonon emission and absorption processes, respectively. The dark-red dots represent the total scattering. (d) Electron scattering rate neat Γ point at 200, 300 and 400 K, represented by green, red and blue dots, respectively.

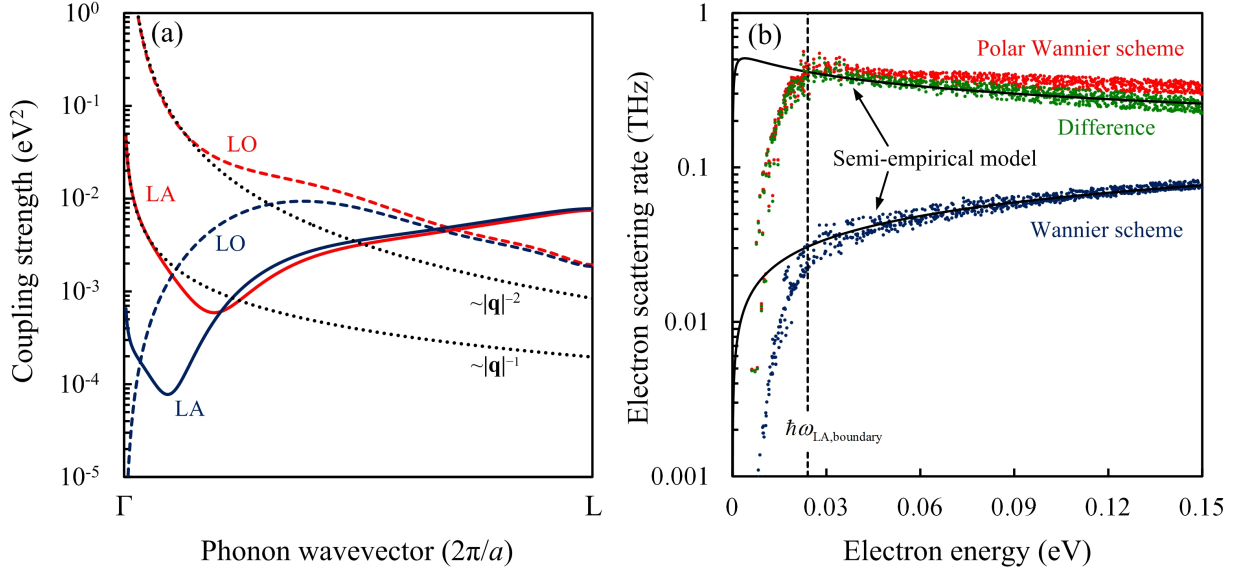

FIG. 3. (a) First-principles coupling strength $|M_q|^2$ between electron at $\Gamma$ point and different phonons in the direction from $\Gamma$ to L points. Red and blue colors denote the using of polar Wannier and Wannier interpolation scheme, respectively. The solid lines are LA phonons and dash lines are LO phonons. The black-dot lines show the $q$ dependences of coupling strength: $|M_q^{PZ}|^2 \sim |q|^{-1}$ and $|M_q^{POP}|^2 \sim |q|^{-2}$ (b) Electron scattering rates due to LA phonons near $\Gamma$ point at 300 K. The red and blue dots are computed by using polar Wannier and Wannier scheme, respectively. The green dots are the difference of scattering rate between the two interpolated schemes. The upper and lower black lines display the trend of scattering rate versus energy of piezoelectric and acoustic-deformation-potential scatterings. It should be emphasized that because the semi-empirical models largely underestimate the acoustic-phonon scattering rates in GaAs, particularly near the $\Gamma$ point, and therefore we have adjusted the prefactors of piezoelectric and acoustic-deformation-potential scatterings listed in table I to fit the simulation results in order to show a better comparison.

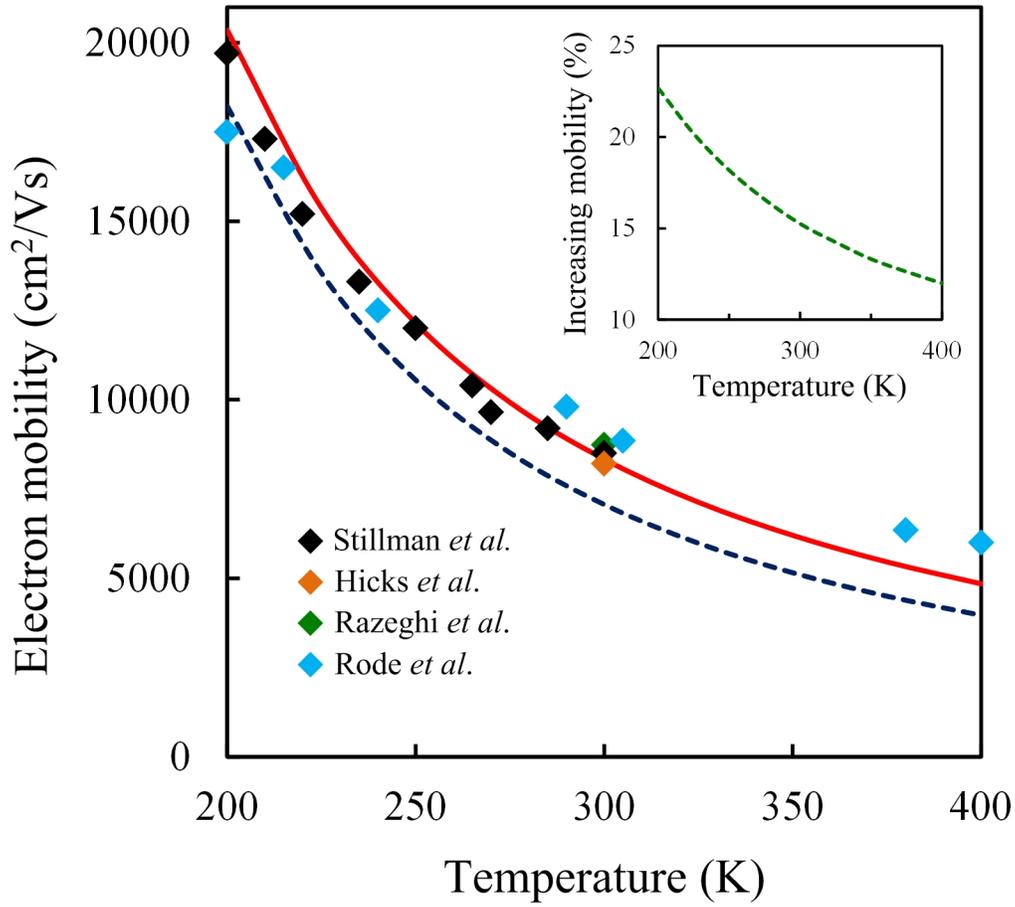

FIG. 4. (a) Electron mobility with respect to temperature from 200 to 400 K. The red and blue-dash lines are the first-principles mobilities computed by iterative scheme and RTA, respectively. The color diamonds are the experimental measurements of the intrinsic GaAs [24,65-67]. The green-dash line in the inset represents the increase of mobility once the LA-phonon scattering is artificially removed.

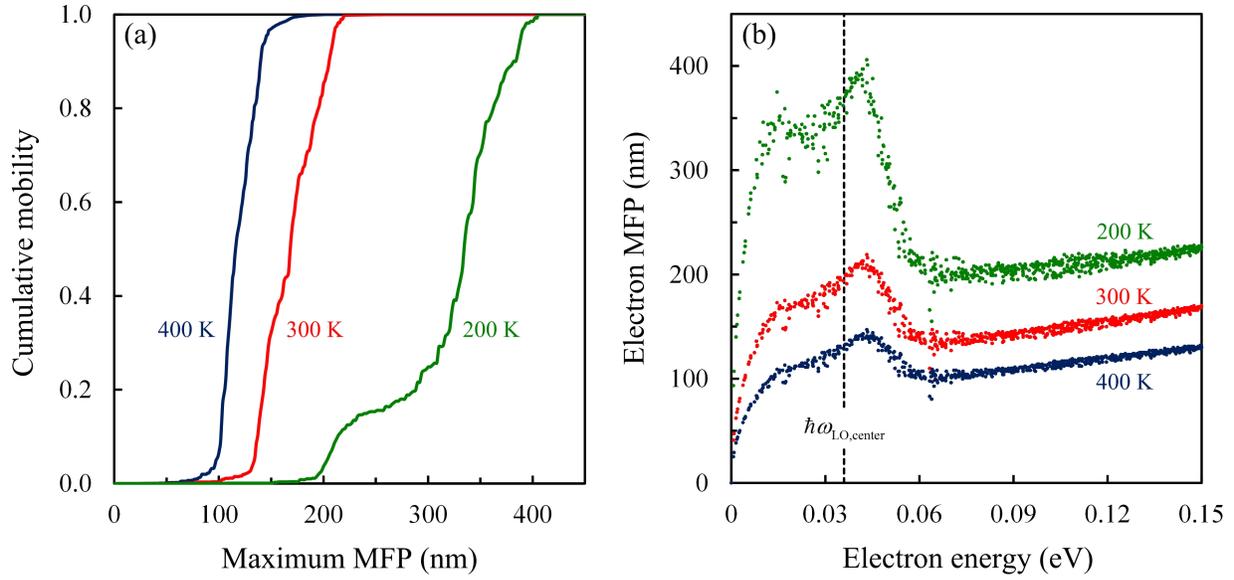

FIG. 5. (a) Cumulative mobilities with respect to different maximum MFPs of GaAs at 200, 300 and 400 K, represented by green, red and blue lines, respectively. The electron MFP spectrum have been normalized by $\mu(\Lambda_{max})$ of each temperature. (b) Electron MFPs near $\Gamma$ point at 200, 300 and 400 K, represented by green, red and blue lines, respectively.

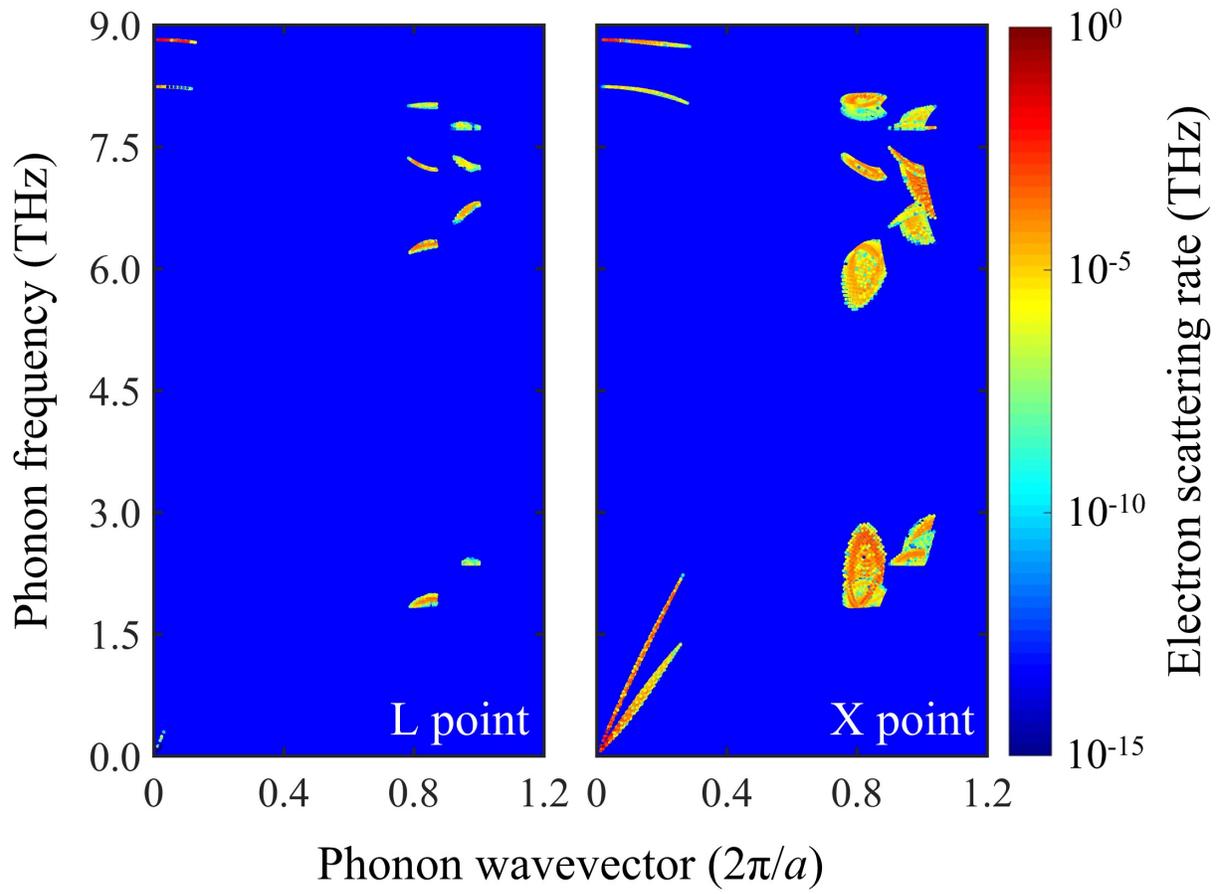

FIG. 6. The electron scattering rates at L (left) and X (right) points due to phonons at 300 K in the entire Brillouin zone, plotted using absolute values of the phonon wavevector.